\documentclass[3p, 10pt, twocolumn]{elsarticle}
\usepackage{graphicx,amscd,amsmath,amssymb,verbatim}
\usepackage{amsfonts,epsfig}
\usepackage{mathptmx}
\usepackage{setspace}
\usepackage{epsfig}
\usepackage{url}
\usepackage{multirow}
\usepackage{subfigure}
\usepackage{color}

\begin{document}

\journal{Elsevier}

\begin{frontmatter}

\title{Investigating the nonequilibrium aspects of long-range Potts model:
refinement of critical temperatures and raw exponents}

\author{Roberto da Silva$^{1}$, J. R. Drugowich de Felicio $^{2}$, Henrique
A. Fernandes$^{3}$}

\address{1-Instituto de F{\'i}sica, Universidade Federal do Rio Grande do Sul, 
Av. Bento Gon{\c{c}}alves, 9500 - CEP 91501-970, Porto Alegre, Rio Grande do Sul, Brazil}

\address{2-Departamento de F\'{\i}sica, Faculdade de Filosofia, Ci\^{e}ncias e Letras de Ribeir\~{a}o Preto, Universidade de S\~{a}o Paulo,\\ 
Av. Bandeirantes, 3900 - CEP 14040-901, Ribeir\~{a}o Preto, S\~{a}o Paulo, Brazil\\}

\address{3-Instituto de Ci{\^e}ncias Exatas, Universidade Federal de Goi{\'a}s, Regional Jata{\'i}, 
BR 364, km 192, 3800 - CEP 75801-615, Jata{\'i}, Goi{\'a}s, Brazil}


\begin{abstract}

In this work, we analyse the $q-$state Potts model with long-range interactions through nonequilibrium scaling relations commonly used when studying short-range systems. We determine the critical temperature via an optimization method for short-time Monte Carlo simulations. The study takes into consideration two different boundary conditions and three different values of range parameters of the couplings. We also present estimates of some critical exponents, named as raw exponents for systems with long-range interactions, which confirm the non-universal character of the model. Finally, we provide some preliminary results addressing the relations between the raw exponents and the exponents obtained for systems with short-range interactions. The results assert that the methods employed in this work are suitable to study the considered model and can easily be adapted to other systems with long-range interactions.


\end{abstract}

\end{frontmatter}

In a famous paper published in 1969, Freeman J. Dyson \cite{Dyson1969}
showed the existence of phase transition in a one-dimensional Ising system
with long-range interactions whose Hamiltonian is given by 
\begin{equation}
\mathcal{H}=-\sum_{i<j}J_{ij}s_{i}s_{j}  \label{Eq:Ising}
\end{equation}%
where $J_{ij}=J(\left\vert i-j\right\vert )$ is the coupling constant and $%
s_{i}=\pm 1$. For that system, two conditions must be fulfilled:%
\begin{equation*}
\begin{array}{lll}
\text{(1)} & I_{1}=\sum_{n=1}^{\infty }J(n)<\infty &  \\ 
&  &  \\ 
\text{(2)} & I_{2}=\sum_{n=1}^{\infty }\frac{\ln (\ln n)}{n^{3}J(n)}<\infty
& 
\end{array}%
\end{equation*}

Such linear interacting spin systems, obviously, do not affect the Landau
argument which refers to local interacting spins. An important class of
long-range (LR) couplings that satisfies such conditions is that of
algebraic decay: 
\begin{equation}
J(n)=\frac{C}{n^{(1+\sigma )}}\   \label{Eq:Coupling}
\end{equation}%
with $0<\sigma <1$, since $\ I_{1}<\int_{1}^{\infty }\frac{dx}{x^{1+\sigma }}%
=\frac{1}{\sigma }$ and 
\begin{equation}
\begin{array}{lll}
I_{2} & \leq & \frac{1}{C}\sum_{n=1}^{\infty }\frac{\ln n}{n^{2-\sigma }} \\ 
&  &  \\ 
& \leq & \frac{1}{C}\int_{1}^{\infty }\frac{\ln x}{x^{2-\sigma }}dx<\infty 
\text{.}%
\end{array}%
\end{equation}%
Here, $\sigma $ is the range parameter of the coupling. Thus, we certainly
expect a phase transition for such $\sigma $-values. This particular and
sufficient condition for the existence of a critical phenomena was
conjectured by Kac and Thompson \cite{Kac1969} in the same year of Dyson's
publication.

The Hamiltonian of the one-dimensional $q-$state Potts model with long-range
interactions can be written as a generalization of the LR Ising model given
by Eq. (\ref{Eq:Ising}) along with the coupling constant presented in Eq. (%
\ref{Eq:Coupling}). It is given by 
\begin{equation}
\beta \mathcal{H}_{\text{Potts}}=-K\sum_{i<j}\ \frac{\delta _{s_{i},s_{j}}}{%
\left\vert i-j\right\vert ^{1+\sigma }}  \label{Eq:Hamiltonian}
\end{equation}%
where $\beta =(k_{B}T)^{-1}$ with $k_{B}$ being the Boltzmann constant and $%
T $ the temperature of the system, $K=\beta J$ is the coupling coefficient, $%
i,j=1,...,L$, and $L$ is the chain length. At equilibrium, one can calculate
the $k$-th moments of magnetization: 
\begin{equation}
\left\langle M^{k}\right\rangle =\frac{1}{L^{k}(q-1)^{k}}\left\langle \left(
\sum_{i=1}^{L}(q\delta _{s_{i},1}-1)\right) ^{k}\right\rangle
\label{Eq:Traditional_order_parameter}
\end{equation}%
where $\left\langle (\cdot )\right\rangle =\frac{1}{Z}\sum_{\left\{
s_{i}\right\} }(\cdot )\exp \left[ -\mathcal{H}_{\text{Potts}}\right] $, and 
$Z=\sum_{\left\{ s_{i}\right\} }\exp \left[ -\mathcal{H}_{\text{Potts}}%
\right] $.

Exactly as reported by Dyson, here it is expected the existence of a
critical temperature $T_{c}$ (or similarly, $K_{c}$). Hence, the
susceptibility $\chi =L\left( \left\langle M^{2}\right\rangle -\left\langle
M\right\rangle ^{2}\right) $, for instance, must behave as $\chi \sim \frac{1%
}{(T-T_{c})}$ at the critical point also for $q>2$ as expected for spin
systems with short-range (SR) interactions.

In 1989, Glumac and Uzelac \cite{Glumac1989} obtained estimates for $K_{c}$
by performing a study for the LR Ising model through a method that scales
the range of interactions. The same authors extended such estimates in 1993
for the LR Potts model by making use of the transfer matrix method \cite%
{Glumac1993}. Although they presented important contributions to the field,
we believe that more attention should be given in systems with LR
interactions by employing new methods and approaches in order to obtain
refined estimates of the critical parameters.

Monte Carlo (MC) equilibrium methods can be an efficient approach to
validate such estimates of $T_c\ (K_{c})$, but in LR systems they are very
expensive. Thus, an interesting alternative to achieve this goal is through
nonequilibrium MC methods. In this context, we can highlight the short-time
dynamics theory deduced and developed by a set of authors from analytical 
\cite{janssen1989} and numerical \cite{Husemaiszheng} points of view. This
approach was developed in the context of model A, according to the
definition of Halperin, Hohenberg, and Ma \cite{Halperin74-77}. This
definition considers the relaxational dynamics of a non-conserved order
parameter described by the solution of the Langevin equation for the
Landau-Ginzburg-Wilson Hamiltonian. Although it has been extensively
investigated for models with SR interactions, systems with LR interactions
have not being subject of study. However, an important aspect of these
interactions is that they can modify the critical equilibrium properties of
the system in consideration.

This method, known as short-time MC simulations, takes into consideration
different time series of the order parameter (the magnetization for most of
the spin models) and its moments. Each time series starts with a fixed
initial magnetization $m_{0}$ and then, the system is quenched from high
temperatures to the critical one. The time evolution of the $k-$th moment of
the magnetization obeys the following general scaling relation: 
\begin{equation}
\overline{M^{k}}(t,\tau ,L,m_{0})=b^{-\frac{k\beta }{\nu }}\overline{M^{k}}%
(b^{-z}t,b^{\frac{1}{\nu }}\tau ,b^{-1}L,b^{x_{0}}m_{0}).
\label{Scaling_relation}
\end{equation}%
Here, $t$ is the time evolution, $b$ is an arbitrary spatial rescaling
factor, $\tau =\left( T-T_{c}\right) /T_{c}$ is the reduced temperature and $%
L$ is the size of the one-dimensional lattice. This evolution is governed by
a new dynamic critical exponent $\theta $ which is independent of the well
known static critical exponents, e.g. $\beta $ and $\nu $, and the dynamic
exponent $z$. This new exponent characterizes the so-called \textit{critical
initial slip}, the anomalous behavior of the magnetization when the system
is quenched to the critical temperature $T_{c}$. In addition, a new critical
exponent $x_{0}$ which represents the anomalous dimension of the initial
magnetization $m_{0}$, is introduced to describe the dependence of the
scaling behavior on the initial conditions. This exponent is related to $%
\theta $ as $x_{0}=\theta z+\beta /\nu $.

Unlike $\left\langle O\right\rangle$, the quantity $\overline{O}$ describes
an average over different random evolutions and initial conditions of the
system. Here, $O$ is a general symbol which means the magnetization or their
superior moments calculated through MC simulations as an average over all $L$
spins and over different $N_{run}$ runs (the number of different time
evolutions): 
\begin{equation}
\overline{O}(t)=\frac{1}{N_{run}\ L}\sum\limits_{r=1}^{N_{run}}\sum%
\limits_{i=1}^{L}O_{i,r}(t),
\end{equation}
where the index $r=1,...,N_{run}$ denotes the corresponding run of each
simulation.

Several authors have performed short-time MC simulations in order to obtain
the following two dynamic critical exponents: the exponent $\theta $, which
governs the critical initial slip of the magnetization $\overline{M}\sim
m_{0}t^{\theta }$ and the exponent $z$ which characterizes the time
correlation in equilibrium (for two good reviews see Albano et al. \cite%
{Albano2011} and B. Zheng \cite{Husemaiszheng}). The exponent $z$, for
instance, can be obtained considering the second moment of the
magnetization, which is written as 
\begin{equation*}
\overline{M_{m_{0}=0}^{2}}=\frac{1}{L^{2d}}\sum\limits_{i=1}^{L^{d}}%
\overline{\sigma _{i}^{2}}+\frac{1}{L^{2d}}\sum\limits_{i\neq j}^{L^{d}}%
\overline{\sigma _{i}\sigma _{j}}\approx L^{-d}
\end{equation*}
for a fixed $t$. By taking into account $k=2$ in Eq. (\ref{Scaling_relation}%
) with $b=t^{1/z}$ and considering that the spin-spin correlation $\overline{%
\sigma _{i}\sigma _{j}}$ is negligible for $m_{0}=0$ (with spins randomly
distributed over the lattice), we obtain the following power law for the
second moment of the magnetization at $T=T_{c}$: 
\begin{equation}
\begin{array}{lll}
\overline{M_{m_{0}=0}^{2}}(t,L) & \approx & t^{\frac{-2\beta }{\nu z}}%
\overline{M_{m_{0}=0}^{2}}(1,t^{-1/z}L) \\ 
\  & \  & \  \\ 
\  & = & t^{\frac{-2\beta }{\nu z}}(t^{-1/z}L)^{-d} \\ 
\  & \  & \  \\ 
\  & \sim & t^{(d-\frac{2\beta }{\nu })/z}\text{. }%
\end{array}
\label{M2}
\end{equation}

Recently, Uzelac \textit{et. al} \cite{Uzelac2008} used the critical
temperatures obtained in their previous works \cite{Glumac1993,Glumac1989}
to perform short-time MC simulations in order to present a preliminary study
of the dynamic critical exponents $\theta $ and $z$ of the Potts model with
LR interactions described by the Hamiltonian given by Eq. (\ref%
{Eq:Hamiltonian}) for the cases $q=2$ and 3.

Although the Ref. \cite{Uzelac2008} shows a lot of interesting things that
motivated this current work, in our opinion, the computing of critical
exponents by simply transposing the finite-size scaling of the short-time
dynamics used for SR systems deserves a lot of further investigations, since
Eq. (\ref{Scaling_relation}) should not work for LR systems, and to the best
of our knowledge, there is nothing in literature suggesting this.

This undoubtedly is not clear as reported by other authors as Chen \textit{%
et. al} \cite{Schulke2000}. That work which has L. Schulke, one of the
precursors in the study of short-time dynamics for SR systems, as one of the
authors, proposes a study of the short-time critical behavior of the $d-$%
dimensional Ginzburg-Landau model with LR interactions. The authors include
an LR term in the Landau-Ginzburg-Wilson Hamiltonian for the time evolution
described by the Langevin equation in order to obtain the short-time scaling
relations similar to that given by Eq. (\ref{Scaling_relation}). The
exponent $\theta $ which characterizes the critical initial slip of the
magnetization is an independent exponent explored in that work. However,
they did not point out a way to explore other power laws to obtain estimates
for other exponents from those LR systems.

Moreover, an important question is if we can use nonequilibrium methods, in
a more fundamental point of view, to estimate critical temperatures of LR
interaction models and not only their critical exponents, since there is not
a consensus about the precision of these estimates from previous results
presented in literature, to the best of our knowledge. So, in this work we
aim to present the estimates of the critical temperatures of the Potts model
with long-range interactions through nonequilibrium methods based on an
optimization method in the context of time dependent Monte Carlo simulations
proposed in Ref. \cite{silva2012}.

In order to answer that question, we look into a simpler power law (for the
initial condition $m_{0}=1$) and keep the traditional order parameter of the
Potts model and their superior moments defined by Eq. (\ref%
{Eq:Traditional_order_parameter}). These ways of analyzing the system were
not considered in Ref. \cite{Uzelac2008} and, as we will show below, they
are important to help shed light on this topic.

Particularly in the case of $(m_{0}=1)$, the system loses the dependence on
initial conditions and the first moment of the magnetization must decay, at
criticality, as 
\begin{equation}
\overline{M_{m_{0}=1}}(t)\sim t^{-\delta }  \label{M1}
\end{equation}%
where $\delta $, which is our first raw exponent for LR systems, is given by 
$\delta =\frac{\beta }{\nu z}$ for models with SR interactions. So, we will
simply denote the exponent by $\delta =\delta _{LR}$ whereas, to the best of
our knowledge, the literature does not show any information about
similarities between short- and long-range exponents.

From now on, we adopt a cautious prescription by considering that the power
laws must exist at the criticality and their exponents are given as raw
exponents. With this assumption, the power law given by Eq. (\ref{M2}) must
be redefined since we do not expect the existence of critical points for
one-dimensional SR systems. Therefore, for LR systems starting with $m_{0}=0$%
, we appropriately consider $\overline{M_{m_{0}=0}^{2}}(t)\sim t^{\xi _{LR}}$%
.

Regarding this letter, our initial intent was to study the localization of
the critical points of the $q-$state Potts Model with LR interactions via
time-dependent MC simulations by estimating the best $K_{c}$ for a given $q$
through a technique based on a statistical concept known as coefficient of
determination. In this approach, we set as input parameter the coupling
coefficient $K^{(\min )}$ (initial value) and run simulations for different
values of $K$ according to a resolution $\Delta K$. In order to show the
robustness of the method, we carried out simulations for $q=2$, $3$, and $4$%
. We also change the range parameter $\sigma $, considering $\sigma =0.7$, $%
0.8$, and $0.9$. With all these analysis in hand, we are just one step to
obtain the critical exponents and explore the universality of the system.
So, we include these estimates as a second part of our study, at the end of
this work. The simulations were carried out by considering free boundary
conditions (FBC). However, we also perform some simulations with periodic
boundary conditions (PBC), in order to compare the results. The latter
boundary condition was considered in Ref. \cite{Dotsenko} for the study of
the LR Potts model through MC simulations at equilibrium. For PBC, the
distance between two sites is $d(i,j)=\min (j-i,i+L-j)$, with $i<j$, such
that $\max_{i,j}d(i,j)=L/2$.

Since at criticality it is expected that the order parameter obeys the power
law behavior given by Eq. (\ref{M1}), we performed MC simulations for each
value $K=K^{(\min )}+i\Delta K$, with $i=1,...,n$, where $n=\left\lfloor
(K^{(\max )}-K^{(\min )})/\Delta K\right\rfloor $, and calculated the
coefficient of determination $r$, which is given by 
\begin{equation}
r=\frac{\sum\limits_{t=t_{\min }}^{t_{\max }}(\overline{\overline{\ln 
\overline{M}}}-a-b\ln t)^{2}}{\sum\limits_{t=t_{\min }}^{t_{\max }}(%
\overline{\overline{\ln \overline{M}}}-\ln \langle M\rangle (t))^{2}},
\label{determination_coefficient}
\end{equation}
with $\overline{\overline{\ln \overline{M}}}=\frac{1}{(t_{\max}-t_{\min })}%
\sum\nolimits_{t=t_{\min }}^{t_{\max }}\ln \overline{M}(t)$. The critical
value $K_{c}$ corresponds to $K^{(opt)}=\arg \max_{K\in \lbrack K^{(\min
)},K^{(\max )}]}\{r\}$ and, $a$ and $b$ are, respectively, the slope and
intercept obtained from the linearization. Here, $t_{\min}$ is the number of
discarded MC steps and $t_{\max}$ the maximum number of MC steps used in our
simulations.

The coefficient $r$ extends from 0 to 1 and has a very simple explanation:
it measures the ratio: (expected variation)/(total variation). So, the
bigger the $r$, the better the linear fit in log-scale, and therefore, the
better the power law which corresponds to the critical parameter excepted
for an order of error $\Delta K$.

\begin{figure}[tbh]
\begin{center}
\includegraphics[width=1.0\columnwidth]{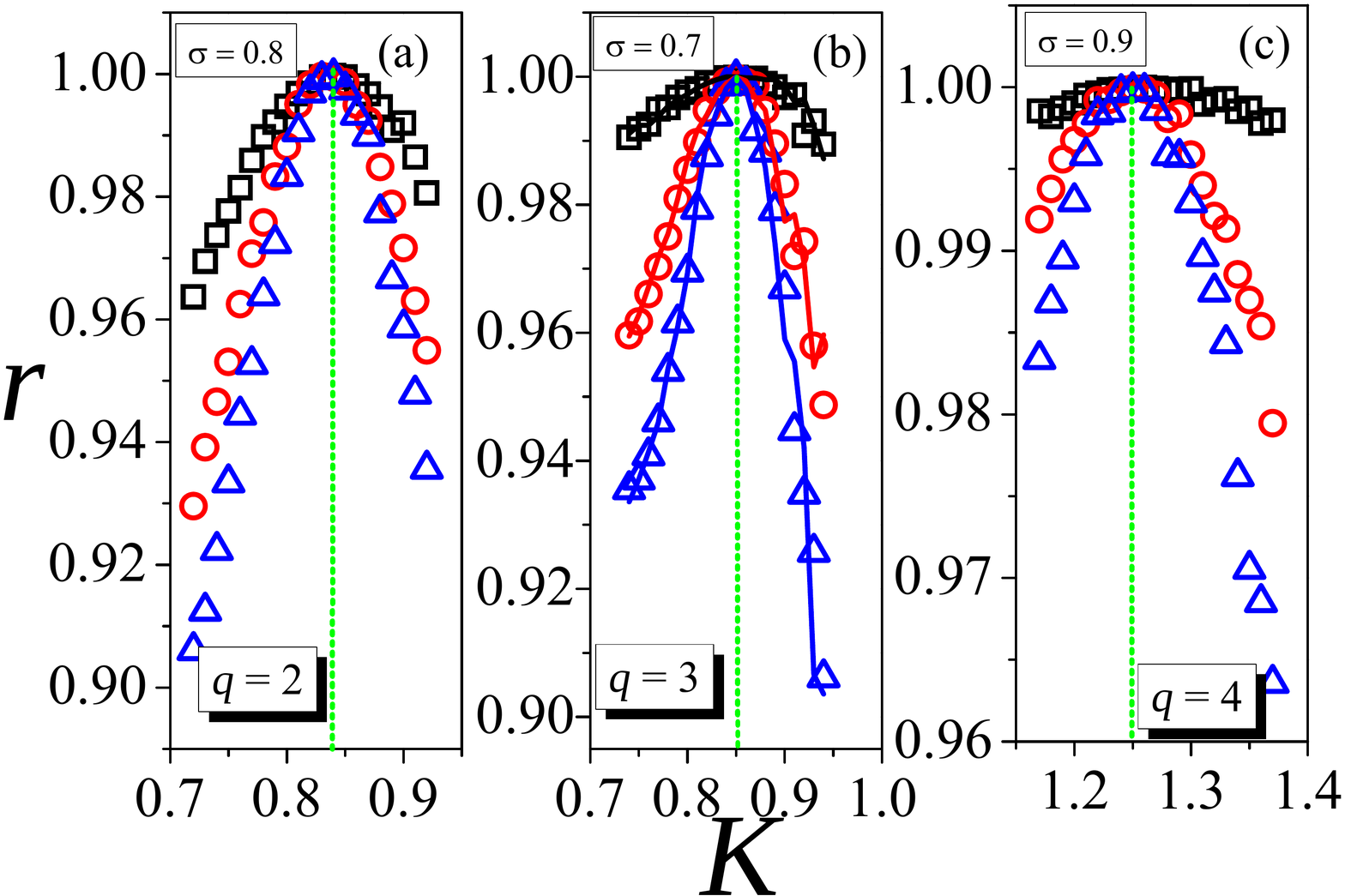}
\end{center}
\caption{Coefficient of determination as function of $K$. We show three
examples: $q=2,\ 3$, and 4, respectively, for three coupling parameters $%
\protect\sigma =0.8$, $0.7$ and $0.9$. Squares, balls, and triangles
correspond to different values of $t_{\max }$ used to calculate the
coefficient of determination: 20, 40, and 60, respectively. The continuous
curves in plot (b) correspond to the results with periodic boundary
conditions.}
\label{Fig:determination}
\end{figure}

Here, we use a very simple procedure: we consider $\Delta K=0.01$, $%
N_{run}=2000$ runs, $L=3000$ sites, and choose (with no previous
information) $K^{(\min )}$ and $K^{(\max )}$ for each value of $q$ and $%
\sigma $. By varying $K$, we are able to determine its optimal value, $%
K^{(opt)}$, which is considered the critical point for the set $(q,\ \sigma)$%
. In our simulations, we used a total number of $N_{MC}=60$ MC steps (where $%
t_{\max}\leq N_{MC}$), which is bigger than that used, for instance, in Ref. 
\cite{Uzelac2008} ($N_{MC}=40$ MC steps). In this approach, we obtain curves
of $r$ as function of $K$ by discarding $t_{\min }=5$ MC steps and varying $%
t_{\max }=20$, 40, and 60, as shown in Fig. \ref{Fig:determination}. 

In this figure, we show three examples of optimizing curves $r\times K$, for 
$q=2,\ 3$, and 4, respectively, for three coupling-parameters $\sigma =0.8,\
0.7$, and $0.9$, for FBC. Figure \ref{Fig:determination} (b) also presents
our estimates for PBC (continuous lines) and, as can be seen, the results
for both boundary conditions are in excellent agreement. So, we can assert
that both FBC and PBC produce good estimates and, therefore, in the
remaining of this work, we consider only FBC to obtain our results.

In order to obtain the final estimates of $K_{c}$, instead of considering a
smaller $\Delta K$ in the region where $r\simeq 1$ to refine our results, we
perform quadratic curve-fittings on $r\times K$ and consider the summit of
each curve as our best estimate. For a comparison of our results with those
ones shown in Ref. \cite{Glumac1993} (based on transfer matrix method), we
decided to present our estimates with four decimal digits. It is important
to notice that sometimes the authors present the results with two, three, or
even four significant digits. We arbitrarily use their estimates with four
significant digits by default.

\begin{figure}[tbh]
\begin{center}
\includegraphics[width=1.0\columnwidth]{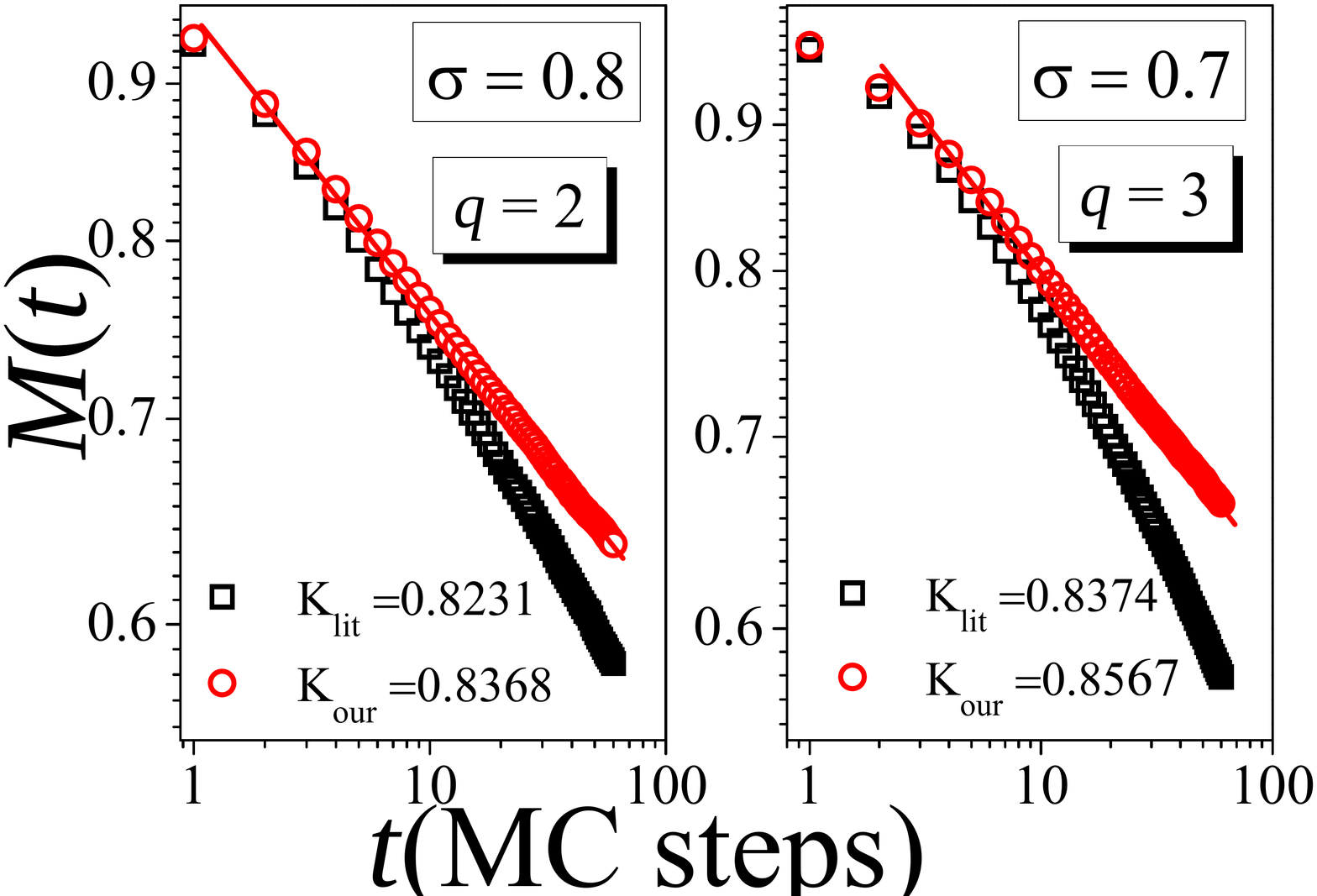}
\end{center}
\caption{Decay of magnetization using our best estimates for $K_{c}$ and
those ones found in Ref. \protect\cite{Glumac1993}. Our results clearly show
better power law behaviors as expected for magnetization at the critical
point when $m_{0}=1$, according to Eq. (\protect\ref{M1}).}
\label{Fig:Literature_vs_us}
\end{figure}

Figure \ref{Fig:Literature_vs_us} shows curves of the time evolution of the
order parameter for the best critical point $K_{c}$ obtained in this work
along with the results found in Ref. \cite{Glumac1993}. 

As can be seen, our results show a notorious visual improvement when
compared with the estimates obtained through equilibrium methods. Although
Fig. \ref{Fig:Literature_vs_us} showed only two cases, the improvement in
results occurs for all set of parameters studied in this work, confirming
the robustness of the methods employed in this work. Our best estimates are
presented in the columns two, four, and six of Table \ref%
{Table:critical_values} and the columns three, five, and seven show the
values obtained from Ref. \cite{Glumac1993}.

\begin{table*}[tbp]
\centering
\begin{tabular}{lllllll}
\hline\hline
$K$ & \multicolumn{2}{c}{$\sigma =0.7$} & \multicolumn{2}{c}{$\sigma =0.8$}
& \multicolumn{2}{c}{$\sigma =0.9$} \\ \cline{2-7}
& This work & Ref. \cite{Glumac1993} & This work & Ref. \cite{Glumac1993} & 
This work & Ref. \cite{Glumac1993} \\ \hline
$q=2$ & $0.7043$ & 0.6833 & $0.8368$ & 0.8231 & $0.9934$ & 0.9973 \\ 
$q=3$ & $0.8567$ & 0.8374 & $0.986\,4$ & 0.9774 & $1.1402$ & 1.1440 \\ 
$q=4$ & $0.957\,6$ & 0.9540 & $1.\,\allowbreak 095\,9$ & 1.0930 & $1.2524$ & 
1.2550 \\ \hline\hline
\end{tabular}%
\caption{Critical coupling coefficients obtained with the method based on
the coefficient of determination for the $q-$state LR Potts model for $q=2$,
3, and 4. We studied these models considering three different range
parameters: $\protect\sigma=0.7$, 0.8, and 0.9. The results are compared
with the estimates found in Ref. \protect\cite{Glumac1993}.}
\label{Table:critical_values}
\end{table*}

With the results of the critical parameters in hand, we decided to look into
the behavior of the power laws related to the magnetization or other more
complex quantities at criticality, as it is traditionally done in the study
of SR systems via time-dependent MC simulations.

For those systems, we showed in Ref. \cite{silva2002a} that combining
simulations with different initial conditions, one produces a cumulant $%
F_{2}(t)=$ $\overline{M_{m_{0}=0}^{2}}(t)/\overline{M_{m_{0}=1}}^{2}(t)$
which, in turn, behaves as $F_{2}(t)\sim t^{d/z}$, where $d$ is the
dimension of the system. So, this cumulant supplies the dynamic critical
exponent $z$ without the need of static exponents previously calculated by
other methods or even conjectured in literature. In this work, we also
conjecture that, for LR systems, a similar behavior is expected, i.e., 
\begin{equation}
F_{2}(t)\sim t^{\gamma }.  \label{F2}
\end{equation}
Here, we would like to reinforce that we do not intent to conjecture any
dependence of the exponents obtained for LR systems with those of SR ones.
For this reason, we present the $\gamma $ as a raw exponent.

In SR models, the static critical exponent $\nu $ can be obtained if the
exponent $z$ is estimated in advance, as for instance, through the cumulant $%
F_{2}$. By using a power law which considers simulations of the order
parameter slightly off the critical temperature $T_{c}\pm \delta $, with $%
\delta <<1$, the derivative $D(t)=\frac{\partial }{\partial T}\ln \overline{M%
}_{m_{0}}(t,T)|_{T=T_{c}}$ can be numerically estimated by $D(t)=\frac{1}{%
2\delta }\ln \frac{\overline{M}_{m_{0}}(t,T_{c}+\delta )}{\overline{M}%
_{m_{0}}(t,T_{c}-\delta )}$. For SR systems, this function behaves as $%
D(t)\sim t^{1/\nu z}$ and, for LR systems, we simply set it as 
\begin{equation}
D(t)\sim t^{\xi }.  \label{D}
\end{equation}

\begin{figure}[tbh]
\begin{center}
\includegraphics[width=1.0\columnwidth]{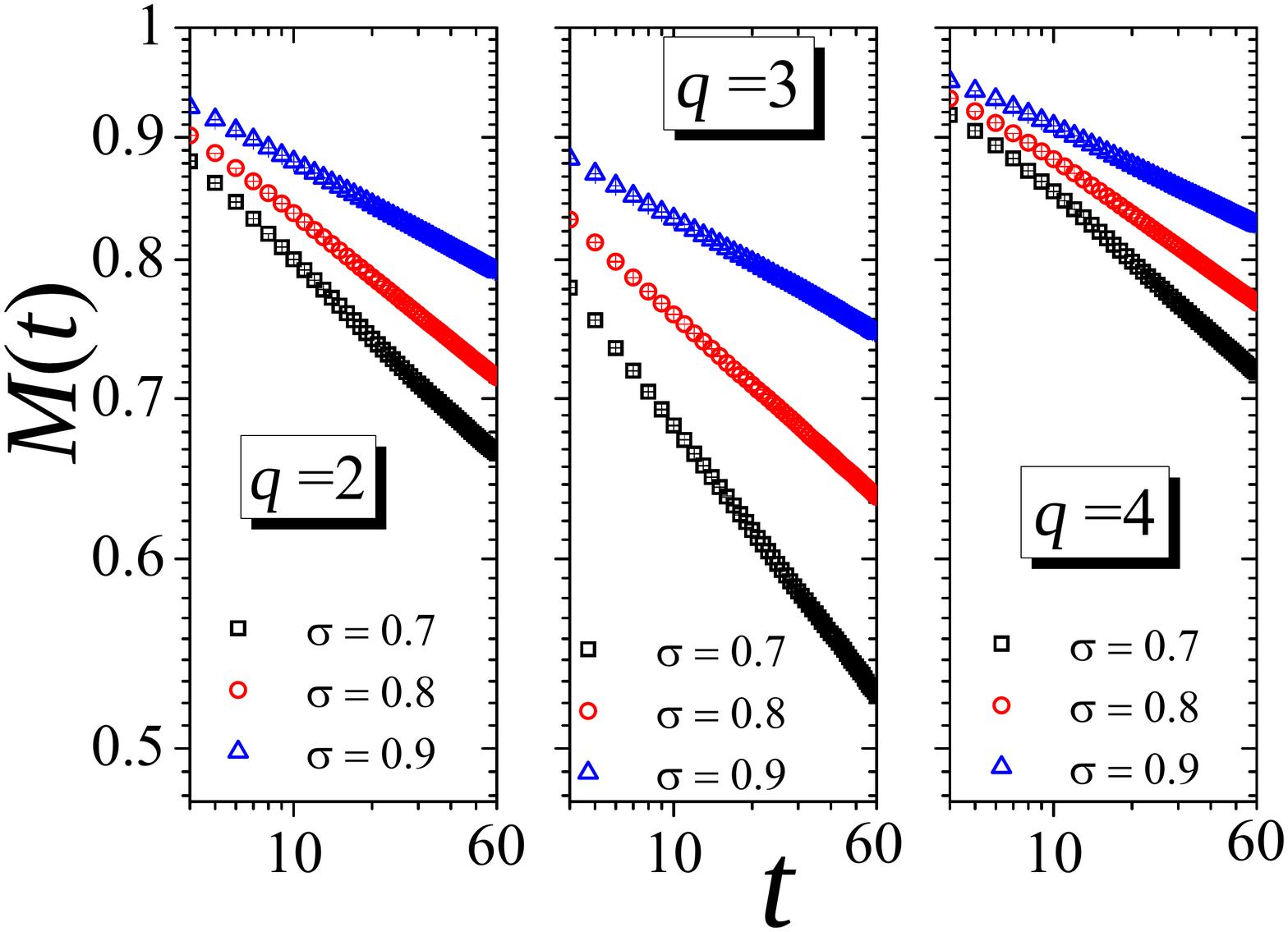} \includegraphics[width=1.0%
\columnwidth]{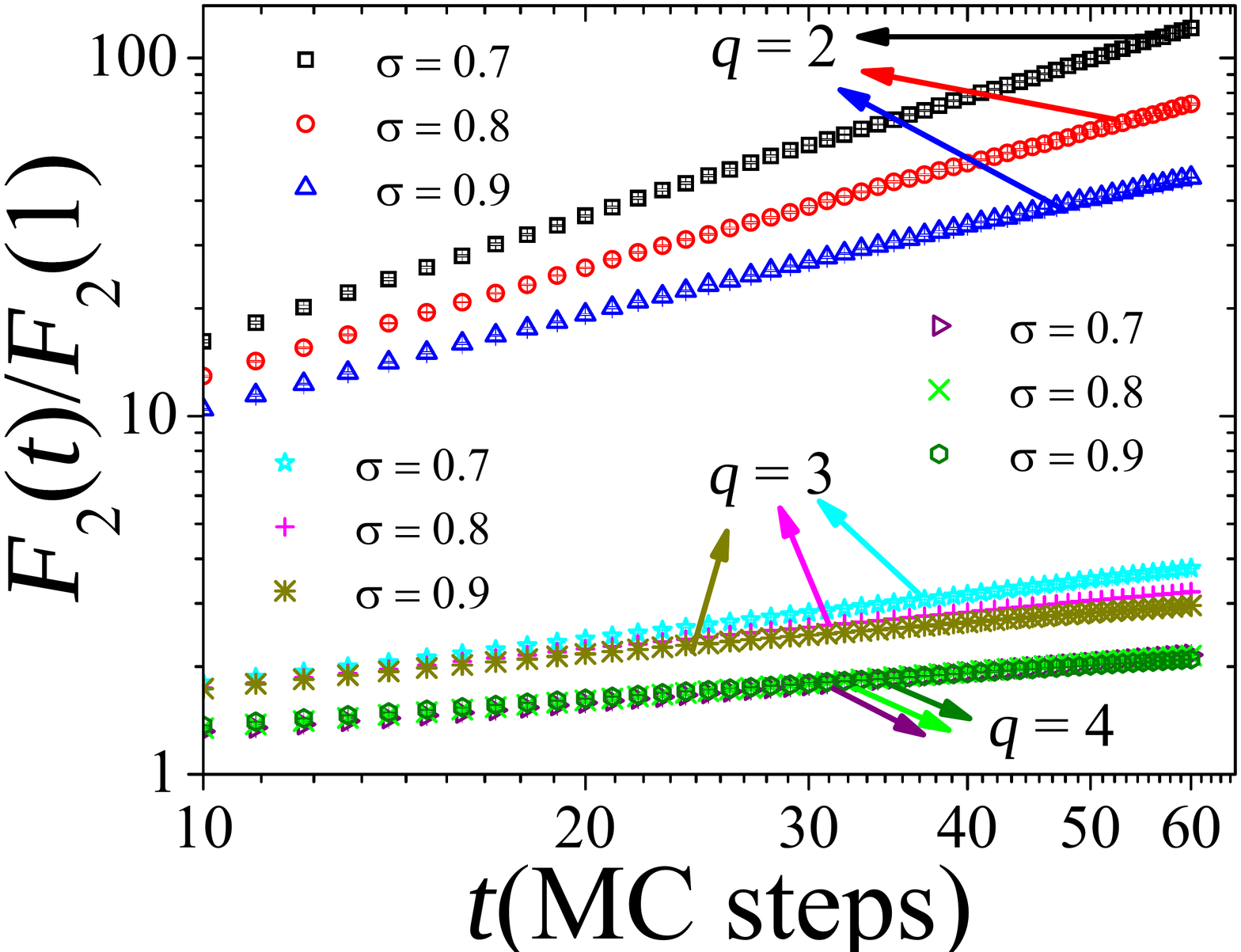} \includegraphics[width=1.0\columnwidth]{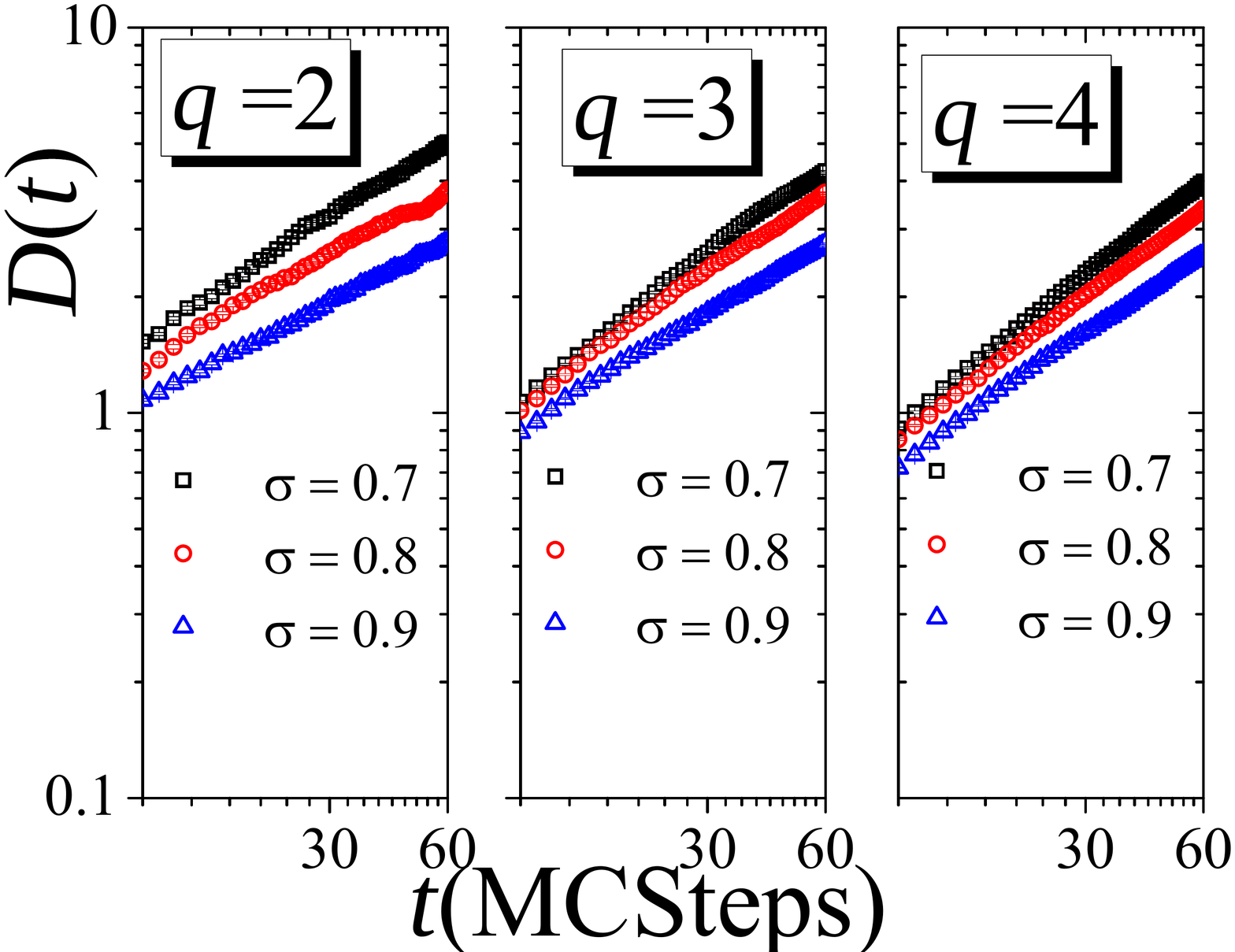}
\end{center}
\caption{Power law behavior of the Eqs. (\protect\ref{M1}), (\protect\ref{F2}%
), and (\protect\ref{D}) for $q=2,\ 3,$ and $4$ and $\protect\sigma =0.7,\
0.8$, and 0.9.}
\label{Fig:power_law_behavior}
\end{figure}

So, from now on we focus our attention on the study of the raw exponents $%
\delta $, $\gamma $, and $\xi $, given respectively by the Eqs. (\ref{M1}), (%
\ref{F2}), and (\ref{D}). In Fig. \ref{Fig:power_law_behavior} we show the
power law behaviors expected for these equations when considering $q=2,\ 3,$
and $4$, and for three values of the range parameter: $\sigma =0.7$, $0.8$,
and $0.9$. These curves were obtained by carrying out simulations of the
model at the best critical parameters obtained above and showed in Table \ref%
{Table:critical_values}. 

From the slope of these power laws (in log scale), we obtained the exponents 
$\delta $, $\gamma $, and $\xi $, and their respective error bars which were
estimated from 5 independent time series averaged over $N_{run}=2000$ runs.
The columns five, nine, and thirteen of Table \ref{Table:Critical_exponents}
correspond to the critical exponents of the standard two-dimensional $q-$%
state Potts model with short-range interactions from Refs. \cite%
{janssen1989,silva2002a,Wu1982}. 
\begin{table*}[tbp] \centering%
\begin{tabular}{l|llll|llll|llll}
\hline\hline
& \multicolumn{4}{l|}{${\tiny q=2}$} & \multicolumn{4}{l|}{${\tiny q=3}$} & 
\multicolumn{4}{l}{${\tiny q=4}$} \\ \hline
${\tiny \sigma }$ & ${\tiny 0.7}$ & ${\tiny 0.8}$ & ${\tiny 0.9}$ & ${\tiny %
d=2}$ & ${\tiny 0.7}$ & ${\tiny 0.8}$ & ${\tiny 0.9}$ & ${\tiny d=2}$ & $%
{\tiny 0.7}$ & ${\tiny 0.8}$ & ${\tiny 0.9}$ & ${\tiny d=2}$ \\ \hline\hline
${\tiny \delta }$ & {\tiny 0.1433(4)} & {\tiny 0.0979(3)} & {\tiny 0.0619(4)}
& {\tiny 0.0580} & {\tiny 0.1017(7)} & {\tiny 0.0881(2)} & {\tiny 0.0589(1)}
& {\tiny 0.0607} & {\tiny 0.0974(3)} & {\tiny 0.0776(1)} & {\tiny 0.0549(1)}
& {\tiny 0.0546} \\ 
${\tiny \gamma }$ & {\tiny 1.100(4)} & {\tiny 0.958(4)} & {\tiny 0.808(5)} & 
{\tiny 0.928} & {\tiny 0.417(1)} & {\tiny 0.339(2)} & {\tiny 0.288(2)} & 
{\tiny 0.911} & {\tiny 0.280(1)} & {\tiny 0.263(1)} & {\tiny 0.231(2)} & 
{\tiny 0.873} \\ 
${\tiny \xi }$ & {\tiny 0.65(2)} & {\tiny 0.58(2)} & {\tiny 0.55(2)} & 
{\tiny 0.46} & {\tiny 0.72(1)} & {\tiny 0.68(1)} & {\tiny 0.59(1)} & {\tiny %
0.55} & {\tiny 0.80(1)} & {\tiny 0.76(1)} & {\tiny 0.68(1)} & {\tiny 0.66}
\\ \hline\hline
\end{tabular}%
\caption{Critical Exponents $\delta$, $\gamma$, and $\xi$, for the $q-$state
Potts model with long-range interactions. There is a tendency of the
exponents to decrease for increasing values of $\sigma$ and $q$, except for
the exponent $\xi$ which increases when $q$ enlarges.}\label%
{Table:Critical_exponents} 
\end{table*}%

As shown in Table \ref{Table:Critical_exponents}, the exponents $\delta $
and $\gamma $ decrease as $q$ enlarges. The same does not occur for $\xi $
which increases for higher values of $q$. However, all exponents decrease
when the interaction exponent $\sigma $ increases, for all values of $q$.
The columns named with $d=2$ present estimates of the exponents calculated
for the two-dimensional SR (each spin on the lattice interacts only with its
nearest neighbors) Potts model (as it is well known, there is no phase
transition for this model when $d=1$). In this case, we consider $\beta $, $%
\nu $, and $z$ obtained in Refs. \cite{silva2002a,Wu1982,Silva2004}. When
comparing our results with the estimates from literature, we observe an
interesting finding: the exponents $\delta $ and $\xi $ approach to the
result for the SR regime as $\sigma $ increases. The most similar case is
for $q=4$ since $\delta _{LR}(\sigma =0.9)=0.0549(1)$ and $\delta
_{SR}=0.0546$, and $\xi _{LR}(\sigma =0.9)=0.68(1)$ and $\xi _{SR}=0.66$.
For a more reliable comparison, we use the same number of significant digits
for the two approaches. We present the SR measures without uncertainty bars
since the only source of error bars is from the exponent $z$ since $\beta $
and $\nu $ are exact measures. In addition, this variation of the exponents
for a given $q$ when $\sigma $ varies confirms that the LR Potts model
exhibits a non-universal behavior.

To explore the only method that should supply the exponent $z$
independently, we can carry out a very preliminary study for the system with
LR interactions by adopting that $\gamma =\frac{d}{z}$ where $d=1$ (which
does not hold when thinking of the SR case). So, by performing this
extrapolation for $q=2$, we obtain $z=0.909(4)$, $1.044(4)$, and $1.238(6)$
for $\sigma =0.7$, $0.8$, and $0.9$, respectively. In Ref. \cite{Uzelac2008}%
, the authors used an alternative order parameter given by $%
M_{x}=L^{-1}q/(q-1)\max_{\alpha }\sum_{i}\left( \delta _{s_{i},\alpha
}-1/q\right) $, denoted by them as absolute value of the magnetization, and
also considered a correspondence between this parameter and the quantity $%
\sqrt{\overline{M_{m_{0}=0}^{2}}}$ deduced according to the scaling relation
valid for SR systems (Eq. \ref{M2}). In that case, they argued that $M_{x}$
behaves as $t^{(d/2-\frac{\beta }{\nu })/z}$ and then considered that $\beta
/\nu =\frac{1-\sigma }{2}$ following Ref. \cite{Brezin1976}. They found $%
z=0.81(1)$, $0.96(4)$ and $1.18(4)$ for $\sigma =0.7$, $0.8$, and $0.9$,
respectively. Although their estimates are similar to our results, there are
some factors which may explain the differences found: 1) We have refined the
critical temperature using short-time Monte Carlo simulations. This value is
used as input in the study of critical exponents and is, therefore, very
important to obtain reliable estimates; 2) We are using a different order
parameter and our method does not use other exponents as input parameters to
obtain the exponent $z$, i.e., by using the cumulant $F_{2}(t)$, the
exponent $z$ is obtained independently.

In our point of view, there are also some issues which must be considered
here:

\begin{enumerate}
\item The power law for $M_{x}$ used in Ref. \cite{Uzelac2008} takes into
consideration only a conjecture and, in addition, some correspondences
deserve much more attention when adapted from SR systems to LR ones. To the
best of our knowledge, the set of static and dynamic critical exponents
presented in the power-laws for systems with SR interactions should not hold
for systems with LR interactions;

\item The authors used $\beta /\nu =(1-\sigma )/2$ from Ref. \cite%
{Brezin1976} as input parameter to obtain the exponent $z$. After a double
check, the Ref. \cite{Brezin1976} presents as estimate, the exponent $\eta
=2-\sigma $. Thus, by considering that $2\beta /\nu =\eta $ (which is valid
only to the two-dimensional SR Potts model), we conclude that $\beta /\nu $
is equal to $1-\sigma /2$ instead of $(1-\sigma )/2$ as used by the authors
in Ref. \cite{Uzelac2008} to obtain $z$ through $M_{x}$. Therefore, we think
that the relation used by them was probably obtained otherwise and, as
pointed out above, we think that it is not valid for LR systems.

\item Finally, if we decided to use the Eq. (\ref{M1}) with $\delta =\beta
/(\nu z)$ and consider both prescriptions for $\beta /\nu $ addressed above
to obtain the exponent $z$, we would find estimates completely different
from those presented in Ref. \cite{Uzelac2008}.
\end{enumerate}

It is important to consider a final comment and some comparisons of our
results with those found in literature. The only case which can be compared
with our results in Ref. \cite{Dotsenko} is for $q=3$ and $\sigma =0.7$. Let
us conjecture that it is possible to relate the well known critical
exponents of SR systems with the raw exponents. In addition, let us suppose
that the exponent $z$ does not appear in the relations for $\delta $ and $%
\xi $. So, if the raw exponents are given by $\delta \rightarrow \delta
_{LR}=\frac{\beta }{\nu }$ and $\xi \rightarrow \xi _{LR}=\frac{1}{\nu }$,
what do we obtain? From our results and these relations, we could estimate $%
\beta $ and $\nu $ independently and compare them with results available in
the literature. Therefore, we performed MC simulations to obtain $\delta
_{LR} $, $\gamma _{LR}$, and $\xi _{LR}$ with both FBC and PBC. The results
for both boundary conditions are in good agreement with each other exactly
as ocurred for the critical parameters presented in Table \ref%
{Table:Critical_exponents}. For PBC, we find $\delta _{LR}=0.102(1)$, $%
\gamma _{LR}=0.419(1)$, and $\xi _{LR}=0.69(1)$ which lead to $\beta =\delta
_{LR}/\xi _{LR}=0.148(3)$ and $\nu =1/\xi _{LR}=1.45(2)$. These results are
surprisingly in absolute agreement with the estimates obtained in Ref. \cite%
{Dotsenko} through the Ferrenberg-Swendsen method \cite{Ferrenberg}: $\beta
=0.15(1)$ and $\nu =1.46(1)$.

It is important to observe that $\beta /\nu =0.102(1)$ does not agree with
the conjecture used by \cite{Uzelac2008} which, for $\sigma =0.7$ is $\beta
/\nu =(1-\sigma )/2=0.15$. Actually, we think that the conjecture must be
true for $q=2$ but not for $q=3$. Let us test this statement by using the
values of Table \ref{Table:Critical_exponents}, this time for $q=2$.
According to our hypothesis that reformulates the exponent to $\delta
_{LR}=\beta /\nu $, we obtain $\beta /\nu \approx 0.143$ for $\sigma =0.7$, $%
\beta /\nu \approx 0.098$, for $\sigma =0.8$, and $\beta /\nu =0.062$ when $%
\sigma =0.9$. If we consider $\beta /\nu =(1-\sigma )/2$, we find, $\beta
/\nu =0.15$, $0.10$, and $0.05$ for $\sigma =0.7$, 0.8, and 0.9,
respectively, which are in fair agreement with our estimates. This shows
that other points deserve a lot of future investigations and the role of $z$
in the raw exponents must be better explored.

In this letter, we presented a useful, suitable, and fast method which has
been successfully used to study systems with short-range interactions, and
now, has proved to be equally efficient when locating critical points of the 
$q-$state Potts model with long-range interactions. This approach, which can
easily be extended to other systems with long-range interactions, allowed us
to obtain the critical temperatures for $q=2$, 3, and 4, and for $\sigma=0.7$%
, 0.8, and 0.9. With these critical parameters in hand, we carried out
short-time Monte Carlo simulations to estimate the exponents $\delta$, $%
\gamma$, and $\xi$, which we call raw exponents. They are related,
respectively, to the power law behaviors of the magnetization $M(t)$, $%
F(t)=M^{2}(t)/[M(t)]^{2}$ when considering different initial conditions, and 
$D(t)=\frac{\partial }{\partial T}\ln M(t,T)|_{T=T_{c}}$. Our results showed
that, for a given $q$, all three raw exponents studied in this work depend
strongly on $\sigma$. This continuous dependence of the critical exponents
on the range parameter $\sigma$ shows that the $q-$state Potts model with
long-range interactions exhibits non-universal behavior.

\textbf{Acknowledgements --} R. da Silva thanks CNPq for financial support
under grant number: 310017/2015-7

\end{document}